\documentclass[twocolumn,fleqn,natbib]{svjour2}
\bibpunct{[}{]}{;}{n}{}{,} 
\smartqed  
\usepackage{graphicx}
\journalname{Granular Matter}

\begin{document}


\title{Solitary wave trains in granular chains 
}

\subtitle{Experiments, theory and simulations}

\titlerunning{Solitary wave trains in granular chains}

\author{St\'ephane Job \and Francisco Melo \and Adam Sokolow \and Surajit Sen}

\institute{ S. Job \at
{\sc Supmeca}, 3 rue Fernand Hainaut 93407 Saint-Ouen, France.
\and F. Melo \at
Departamento de F\'{\i}sica, USACH and CIMAT,\\
Av. Ecuador 3493, Casilla 307, Correo 2, Santiago de Chile.
\and A. Sokolow and S. Sen \at
Department of Physics, State University of New\\
York at Buffalo, Buffalo, New York 14260, USA.}

\date{Received: date}

\maketitle

\begin{abstract}
The features of solitary waves observed in horizontal monodisperse chain of barely touching beads not only depend on geometrical and material properties of the beads but also on the initial perturbation provided at the edge of the chain. An impact of a large striker on a monodisperse chain, and similarly a sharp decrease of bead radius in a stepped chain, generates a solitary wave train containing many single solitary waves ordered by decreasing amplitudes. We find, by simple analytical arguments, that the unloading of compression force at the chain edge has a nearly exponential decrease. The characteristic time is mainly a function involving the grains' masses and the striker mass. Numerical calculations and experiments corroborate these findings.
\keywords{one-dimensional granular chain \and solitary wave \and stepped chain \and pulses train formation}
\end{abstract}

\section{Introduction}\label{sec:introduction}

The problem of mechanical energy transport in an alignment of grains that is held within boundaries has attracted significant attention in recent years~\cite{Nesterenko1983,Lazaridi1985,Sinkovits95,Sen96,Coste1997,Sen98}. It is well established that sound waves may propagate in such a medium and acoustic speed increases with imposed static stress. In the presence of loading, linear acoustic waves disperse as a function of time and space~\cite{Nesterenko2001,Sen2001b}. When the chain is not loaded, i.e., when the grains barely touch one another, acoustic speed vanishes: the medium cannot sustain linear acoustic waves. In this {\em sonic vacuum} limit, nonlinear phenomena predominate and any sharp enough perturbation is known to result only in a solitary wave which travels through the chain~\cite{Nesterenko1983}.

It is well known that mechanical energy transport from one grain to another, when the grains are barely touching, is a highly nonlinear process. The spherical grains interact via the intrinsically nonlinear Hertz potential, in which the repulsive potential between the grains increases with grain-grain overlap via a $5/2$ power law. This power law is softer than the quadratic (harmonic) repulsion for small overlaps but rapidly becomes steeper as compression increases. Hence, it is energetically expensive for two adjacent grains to sustain contact for long. This physical feature is at the heart of the peculiar nonlinear properties associated with mechanical energy transport in granular alignments. It can be shown that whenever the repulsive potential has a steeper than quadratic power law growth in the overlap (or intergrain distance), mechanical energy transfer from one grain to the next starts off slowly and then ends abruptly. Energy is hence transported as a ``lump.'' These alignments therefore propagate energy in the form of solitary waves. From a geometrical standpoint, the solitary wave in a granular alignment of spheres is a highly symmetric object with a well defined center and its velocity depends upon its amplitude or on the amount of energy it carries. Details of the properties of these solitary waves can be found elsewhere~\cite{Nesterenko1983,Coste1997,Nesterenko2001,Sen2001b,Job2005}.

Attention has now turned to mechanical energy propagation due to initial perturbations that generate significant grain compressions at a chain end. When kinetic energy is given to a single grain at time $t=0$, grain-grain interactions force the system to partition this energy into part kinetic and part potential energies. The energy partitioning between kinetic and potential is controlled by the nature of the intergrain potential, in this case, the Hertz potential. The kinetic energy imparted to each edge sphere hence naturally matures into a single solitary wave in space and time. When the contact time of an impact on the edge grains in a non-loaded alignment becomes sufficiently large, acoustic waves propagation facilitates to rapidly distribute the grain compression among multiple contacts. Thus, longer-lived contact times at a chain edge can generate comparable grain-grain overlaps in several contacts.

Lazaridi and Nesterenko were the first to demonstrate that an initial perturbation provided at the edge of a non-loaded granular chain by a large enough striker can result in the formation of solitary wave trains, containing many single solitary waves ordered by decreasing amplitudes~\cite{Lazaridi1985}. These observations have been confirmed and extended in a recent simulational work~\cite{Sokolow2007}. The force amplitude of the $p^{th}$ single pulse in a solitary wave train has been found to decrease approximately exponentially, i.e. $\max{[F(p)]}\sim\exp{(-\alpha p)}$. The exponential coefficient $\alpha$ was demonstrated to depend only on striker/bead mass ratio, even though the attempt to find a simple relationship failed~\cite{Sokolow2007}. In an alternate way of generating solitary wave trains, it has also been shown~\cite{Sokolow2007} that if $N_0$ edge grains are simultaneously set moving from their equilibrium positions in the direction of the chain at some initial time $t=0$, then $N_0$ independent solitary waves of progressively diminishing amplitudes form in the chain. That's exactly what one may expect in a Newton's pendulum with several spheres: when two or more spheres are used to initiate a swing, just as many spheres swing out at the far end~\cite{Herrmann1981,Herrmann1982,Nesterenko1983}. These numerical findings also agree with observations in a vertical column of grains colliding with a rigid wall~\cite{Falcon1998}, for which the time of collision increases linearly with the number of grains contained in the chain due to energy redistribution of intrinsically dispersive compression waves. In another way, several authors have also demonstrated solitary wave trains may form in granular systems containing soft and hard materials~\cite{Job2005,Nesterenko2005,Daraio2006a,Daraio2006b}.

In this work, we report experimental studies, detailed analysis, and numerical simulations, to understand how the nature of the initial perturbation provided to the edge of a non-loaded granular chain can result in the subsequent formation of solitary wave trains. The experimental setup under study is a {\em stepped chain}, defined as two adjacent and non-loaded monodisperse granular alignments. A stepped chain is similar, from the point of view of the interaction of the last bead of the first chain with edge beads of the second smaller chain, to a monodisperse chain impacted by a large and massive striker~\cite{Lazaridi1985,Nesterenko1994,Nesterenko1995,Nesterenko2005}. A stepped chain may also be seen as a particular case of a tapered chain (i.e., a chain with progressively decreasing bead radius)~\cite{Sen2001a,SenNaka2005,Doney2005,Melo2006,Doney2006}. Nesterenko {\it et al}, then considering a system similar to a stepped chain to study the interaction between two {\em sonic vacua}, reported total energy and momentum transmissions when a pulse passes through a sharp decrease of bead's size, and partial reflection in the reverse case~\cite{Nesterenko1994,Nesterenko1995}. Such nontrivial and fascinating behaviors are highly related to the way energy and momentum transfer nonlinearly at the interface between chains. The aim of our work is to determine the features of the solitary wave train generated when an impulse, initiated at the larger edge beads, crosses the radius mismatch. Starting from the conclusions derived in~\cite{Sokolow2007}, we provide here a detailed relationship between the shape of the solitary wave train and the bead's mass ratio. Making use of the similarity mentioned above between the two chain types, our simulational studies model a monodisperse chain with large striker mass.

In section~\ref{sec:apparatus}, we present the experimental apparatus and recall the main elements of a descriptive model based on long wavelength approximation of nonlinear wave propagation. In section~\ref{sec:qualitative}, we discuss the experiments, by means of a qualitative analysis. Finally, in section~\ref{sec:robust}, we construct a complementary quantitative description and compare it with numerical simulations.

\section{Experimental apparatus and long wavelength description of nonlinear wave propagation}\label{sec:apparatus}

\begin{figure}[top]
\centering\includegraphics{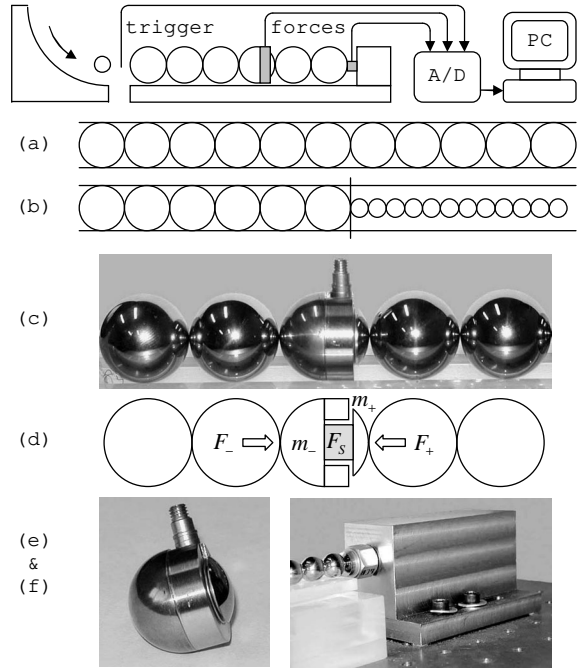}
\caption{Experimental setup: (top) Schematic view showing a chain with sensors and data acquisition facilities. (a) Monodisperse chain and (b) stepped chain. (c-d) Portion of the chain containing a force sensor embedded in a bead (e) and ended by rigid and non moving force sensor (f).}\label{fig:fig1}
\end{figure}

The setup under experimental study is a chain with a steep decrease of bead radius, namely a {\em stepped chain}. All the beads are made of same material (density $\rho$, Young modulus $Y$, and Poisson ratio $\nu$), but their radii $R_n$ and masses $m=(4/3)\pi\rho R_n^3$ differ from one chain to the other: $R_1$ in the first part and $R_2\neq R_1$ in the second one. The stepped chain is impacted by a spherical striker initially moving at velocity $V_s$ and whose radius and mass are $R_s$ and $m_s$. In our experiment the beads and the striker are high carbon chrome hardened {\it AISI 52100} steel roll bearings with $Y=203$~GPa, $\nu=0.3$, and $\rho=7780$~kg/m$^3$. Typical elements of our experimental setup are shown in Fig.~\ref{fig:fig1}. A nonlinear wave is initiated in the chain by the striker impact and measurements are achieved by using two piezoelectric force sensors. One of the sensors is boxed in a rigid and non moving heavy piece of metal placed at the end of the chain, and the second one is inserted inside a $13$~mm radius bead that has been cut in two parts. The mass of the bead-sensor element has been rectified to match the mass of an original bead~\cite{Job2005}, and the force measured by the embedded sensor is $F_s\simeq(m_{-}/m)F_{+}+(m_{+}/m)F_{-}$, where $F_{\pm}$ (resp. $m_{\pm}$) refer to front and back forces (resp. masses) of the bead-sensor element~\cite{Job2005}. Assuming the mass in front of the embedded sensor to be negligible ($m_{+}\ll m$ and actually $m_{+}/m\simeq0.11$) allows to have a satisfactory estimation of the force exactly at the contact with neighbor bead.

In accordance with pioneering works of Nesterenko~\cite{Nesterenko1983,Lazaridi1985,Nesterenko2001}, the striker impact generates an elastic deformation of contact regions between beads, which propagates from one to another. The overlap at contact $n$ (i.e. between beads $n$ and $n-1$) is defined as $\delta_n=(u_{n-1}-u_{n})$, where $u_{n}$ is the position of bead number $n$. Derived from Hertz potential~\cite{Hertz1881}, the force at contact $n$ is $F_n=\kappa\delta_n^{3/2}$ if $\delta_n>0$ (and $0$ if $\delta_n\leq0$), where $\kappa=R^{1/2}/2^{3/2}\theta$ and $\theta=3(1-\nu^2)/4Y$ are constants. More precisely, the dynamics of the chain is described by applying Newton's second law to each of the beads: $m\ddot{u}_n=(F_{n}-F_{n+1})$. A continuum limit of this set of equations can be derived under the long wavelength approximation, when characteristic wavelength $\lambda$ of the deformation is greater than the radii of the beads~\cite{Nesterenko2001}. Keeping terms of up to the fourth order of small parameter $(2R/\lambda)$ in Taylor expansions of $u_{n\pm1}=u(x_n\pm2R,t)$ versus $u_{n}=u(x_n,t)$ leads to a nonlinear equation for the strain $\psi(x,t)=-\partial_xu(x,t)$,
\begin{equation}\label{eq:strain_equation}
\ddot{\Psi}\simeq (2R)^{5/2}(\kappa/m)[\Psi^{3/2}+(2/5)R^2\Psi^{1/4}(\Psi^{5/4})_{xx}]_{xx}.
\end{equation}
An exact solution of the last equation is,
\begin{equation}\label{eq:strain_solution}
\Psi=\Psi_m\cos^4{\xi},\ \xi=\frac{x-vt}{R\sqrt{10}},\ 
v=(6/5\pi\rho\theta)^{1/2}\Psi_m^{1/4},
\end{equation}
where one hump of the strain wave $\Psi$, for $|\xi|\leq\pi/2$, describes a single solitary wave (SW) solution, with a compact spatial extent of the order of few times the characteristic length $\lambda=R\sqrt{10}$. Thereby, the order of magnitude of the neglected terms, $(2R/\lambda)^5\simeq10\%$, provides an estimate of the accuracy of the continuous solution, Eq.~\ref{eq:strain_solution}. To complete the long wavelength description, one can estimate overlap and force at contact $n$ and velocity of bead $n$ from the continuous wave description, as $\delta_n\simeq2R\times\Psi[x=(2n+1)R,t]$, $F_n=\kappa\delta_n^{3/2}$, and $V_n\simeq v\times\Psi(x=2Rn,t)$.

\section{Experimental observations and qualitative analysis under weak interaction assumption}\label{sec:qualitative}

\begin{figure}[top]
\centering\includegraphics{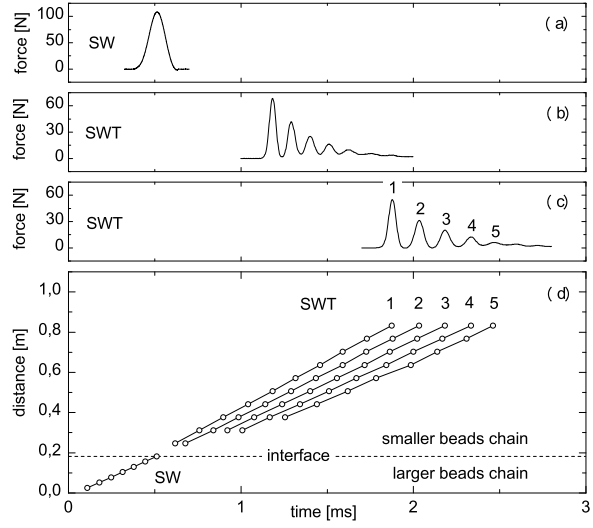}
\caption{Experiments in stepped chain made with $7$ beads of radius $R_1=13$~mm followed by up to $50$ beads of radius $R_2=6.5$~mm. Force is measured at the end of the chain, and the number of beads is progressively increased to follow pulse modification as it propagates deeper in the stepped chain. (a) Force as a function of time, describing a SW measured at the end of a monodisperse chain of $7$ beads of radius $R_1$. This SW then generates a SWT when entering the chain of beads with radius $R_2$. (b-c) SWT in the second part of the stepped chain, respectively at intermediate position ($7$ large beads followed by $25$ smaller beads) and at the end of the stepped chain ($7$ large beads followed by $50$ smaller beads). (d) Position of local maxima of the force as a function of their respective time of flights. Wave splitting after passing through the interface is clearly shown.}\label{fig:fig2}
\end{figure}

We present experimental observations carried out first in a monodisperse chain, and second in a stepped chain. The aim of this section is to derive qualitative estimates based on the crude assumption that bead interactions are described through weakly coupled collisions. We will also assume that the interaction between the two parts of the stepped chains is governed by the interaction between edge beads of each chains. Thus, the last bead of the first chain is considered as a striker for the second chain with smaller beads.

As previously demonstrated experimentally and numerically~\cite{Coste1997,Chatterjee1999,Sen2001b,Job2005}, if the initial impact is applied by a light striker ($m_s\ll m$), a single solitary wave (SW) propagates in the chain. Such a wave is shown in inset $(a)$ of Fig.~\ref{fig:fig2}, which represents the force measured at the end of a short monodisperse chain containing $7$ beads of radius $R_1=13$~mm, and resulting from the impact of a striker of radius $R_s=6.5$~mm. The profile and the velocity both exhibits satisfactory agreements with theoretical counterparts presented in Eq.~\ref{eq:strain_solution}~\cite{Lazaridi1985,Coste1997,Nesterenko2001,Job2005}.

A necessary condition for the observation of a single SW should be the striker rebound after a single collision: the striker won't ever impact the chain again. Applying energy and momentum conservation for describing the initial striker impact on the first bead of the chain (considered uncoupled from the rest of the chain), striker velocity $V_s$ and first bead velocity $V_f$ become after their collision,
\begin{equation}\label{eq:velocity_ratio}
V'_s=\left(\frac{m_s-m}{m_s+m}\right)V_s\mbox{ and }V'_f=\left(\frac{2m_s}{m_s+m}\right)V_s,
\end{equation}
which indicates that striker velocity is negative after impact if it is lighter than a bead ($m_s<m$). Then, the first bead of the chain accelerates right after striker impact and subsequently transmits energy and momentum to next nearest neighbor through an almost perfectly nonreflecting collision. From one bead to another, this process might generate a wave with compact spatial support in the chain, i.e. a single SW. Based on this qualitative description, the amplitude of this wave, expressed in terms of bead velocity, might thus be roughly,
\begin{equation}\label{eq:velocity_amplitude}
V_m\sim\left(\frac{2m_s}{m_s+m}\right)V_s.
\end{equation}

Another condition for the observation of a single SW, should also be an initial collision duration shorter or equal to the duration of the transmitted SW, which is approximately two times the characteristic time $\tau_{sw}=R\sqrt{10}/v$. Hence, fraction of initial energy and momentum will be transmitted through a duration shorter than the characteristic time of the chain. The duration of the collision between the striker and the first bead (considered uncoupled from the rest of the chain) is $\tau_{c}=(2.94/V_s^{1/5})(\mu/k)^{2/5}$, where $\mu=1/(m^{-1}+m_s^{-1})$ is the reduced mass and $k^{-1}=(5\theta/2)(R^{-1}+R_s^{-1})^{1/2}$~\cite{Landau1967}. Using first, the relation between $v$, $\Psi_m$ and $V_m$, second the relation between $V_m$ and $V_s$ given by Eq.~\ref{eq:velocity_amplitude}, and finally introducing the radius ratio $\eta=R/R_s$, the ratio of durations,
\begin{equation}\label{eq:duration_ratio}
\frac{\tau_{c}}{2\tau_{sw}}
=\frac{2.94}{\sqrt{10}}\left[\frac{1+\eta}{(1+\eta^3)^2}\frac{V_m}{2V_s}\right]^\frac{1}{5}
\sim \left[\frac{1+\eta}{(1+\eta^3)^3}\right]^\frac{1}{5},
\end{equation}
demonstrates that $\tau_{c}\leq2\tau_{sw}$ if the striker is lighter than a bead ($m_s<m$ i.e. $\eta\geq 1$).

On the other hand, if the striker is heavier than the beads, the impact of the striker leads to the generation of a solitary wave train (SWT)~\cite{Nesterenko1983,Lazaridi1985,Nesterenko2001,Sokolow2007}. Such a SWT, resulting in our experiment from the transmission of a single SW from the first part of the chain to the second part with smaller beads, is shown in insets $(b)$ and $(c)$ on Fig.~\ref{fig:fig2}. These insets represent the force measured at the end of a stepped chain containing $7$ beads of radius $R_1=13$~mm followed respectively by $25$ and $50$ beads of radius $R_2=6.5$~mm. One sees clearly that the incident SW, shown in inset $(a)$ on Fig.~\ref{fig:fig2}, splits into a SWT right after passing through the interface between the chains. This SWT, considered far from the interface, contains many SWs ordered by decreasing amplitudes and behaving individually according to theoretical predictions of Eq.~\ref{eq:strain_solution}. The wave splitting is also clearly underlined in inset $(d)$ of Fig.~\ref{fig:fig2}, which represents the position of few noticeable local maxima in the force signal (i.e. the position where force was measured) as a function of their respective time of flight. Finally, analyzing the local maxima of the force far from the interface shows that the force amplitude of the $k^{th}$ SW in the SWT behaves like $F_m(k)\propto\exp{(-\alpha k)}$. Matching experimental data with an exponential fit, at nine different positions in the second part of the chain, one finds $\alpha_{exp}=0.48\pm0.03$.

\begin{figure}[top]
\centering\includegraphics{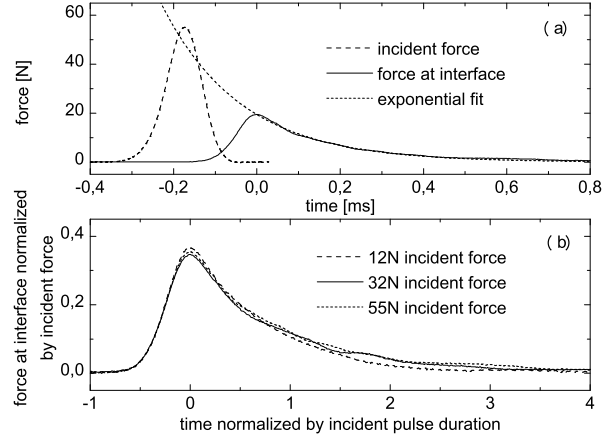}
\caption{Experiments in stepped chain: (a) force versus time (incident and interface). One sees first a characteristic compression of the incident solitary wave followed by longer exponential relaxation. (b) Forces at interface normalized by incident force amplitude versus time normalized by duration of incident wave. Matching experimental data to an exponential fit for each of the force amplitudes indicates that unloading characteristic time is such that $\tau/2\tau_{sw(1)}=0.86\pm0.08$.}\label{fig:fig3}
\end{figure}

A closer view of what happens at the interface is made possible by using the embedded sensor to measure the force as a function of time exactly at the interface between the chains. This measurement is presented in inset $(a)$ of Fig.~\ref{fig:fig3}, which represents the incident force (the force measured at the contact before the interface) and the force at the interface. One sees that the interface is first loaded within a short duration and is then unloaded through a relaxation-like process which roughly follows an exponential decrease, with characteristic time much larger than the loading phase. To obtain a deep insight of this process, the force at the interface, measured for different incident force amplitudes, and normalized by the incident force amplitude is plotted in inset $(b)$ of Fig.~\ref{fig:fig3} as a function of time normalized by the characteristic time of the incident force. This plot shows a collapse of the data in a large range of the incident force amplitude. This demonstrates, first, that the loading duration of the interface equals the loading time of the incident wave (i.e. half its duration). Second, the maximum force at the interface $F_m^{(i)}$ is proportional to the incident SW's force amplitude $F_m^{(1)}$. Third, the unloading characteristic time $\tau$ of the interface is proportional to the incident SW's duration $2\tau_{sw(1)}$. In the experimental range of incident forces, we find $F_m^{(i)}/F_m^{(1)}=0.36\pm0.01$ and matching the force at interface by an exponential fit indicates $\tau/2\tau_{sw(1)}=0.86\pm0.08$.

As can be seen in Fig.~\ref{fig:fig2} and Fig.~\ref{fig:fig3}, the total duration of the compression at the interface and of the SWT far from the interface is much greater than the duration of the incident single SW. This result may be less obvious since, from Eq.~\ref{eq:duration_ratio}, the duration of the initial collision is of the order of one single transmitted SW duration ($\tau_{c}\simeq2\tau_{sw}$ if $R_s\gg R$). In fact, following our qualitative description, this suggests that few secondary collisions between a large striker (i.e. the last bead of the first chain) and the first bead of the second chain may occur. This assumption agrees with Nesterenko's numerical simulations, demonstrating that a sequence of pulses is induced at the interface of two sonic vacua by the deceleration of interfacial particles~\cite{Nesterenko2005}. Indeed, one can note from Eq.~\ref{eq:velocity_ratio} (with $m_s>m$) that the striker is decelerated after the initial collision but continues to move forward, while the first bead of the chain is accelerated to a higher velocity ($V'_f>V'_s$). The striker and the first bead thus tend to separate due to velocity differences, and stress and strain at their contact quickly vanishes. Almost entire transfers of momentum and energy from the first bead to next nearest neighbor occur during this process and the first bead consequently decelerates while generating a single SW with an amplitude $V_m(1)$ given by Eq.~\ref{eq:velocity_amplitude}. The amplitude of the first SW, expressed in terms of compression force of bead contacts, is such that $F_m(1)\propto V_m^{6/5}(1)$. The first bead thus ends up being decelerated, an increase of stress and strain applied by the striker consequently occurs, namely a secondary collision. The velocity of the first bead after this second collision is such that $V_m(2)=(m_s-m)/(m_s+m)V_m(1)$ (see Eq.~\ref{eq:velocity_ratio}). Following this simple picture, the duration between two successive secondary collisions should be of the order of the duration of one single transmitted SW. The repetition of these secondary collisions (acceleration and deceleration of the first bead) leads to the generation of successive single SWs whose amplitudes decrease exponentially. Force amplitude of the $k^{th}$ transmitted SW is thus $F_m(k)\propto[(m_s-m)/(m_s+m)]^{6k/5}\propto\exp{(-\alpha k)}$ with,
\begin{equation}\label{eq:alpha_simple}
\alpha\sim\left(\frac{6}{5}\right)\ln{\left(\frac{m_s+m}{m_s-m}\right)}\mbox{ if }m_s\geq m.
\end{equation}

Interestingly, this qualitative description of repeated secondary collisions predicts the total transmission of striker energy and momentum to the second chain, as observed by Nesterenko {\it et al}~\cite{Lazaridi1985,Nesterenko1994,Nesterenko1995}, over an infinite duration.

In addition, denoting $t_k$ as the instant of the $k^{th}$ secondary collision of the striker with the first bead and introducing $\tau_*=(t_{k}-t_{k-1})$ as the elapsed duration between two successive secondary collisions of the striker, and finally considering that the force at the interface decreases approximately exponentially within a characteristic duration $\tau$, one can predict that the unloading of the interfacial force versus time is such that,
\begin{equation}\label{eq:tau_simple}
F_{int}(t_k)\propto\exp{(-t_k/\tau)}\mbox{ with }\tau\simeq\tau_{*}/\alpha.
\end{equation}

The elapsed time $\tau_*$ between two successive secondary collisions of the striker can be estimated from the experimental measurements of $\tau/2\tau_{sw(1)}$ and $\alpha_{exp}$, such that $\tau_*/2\tau_{sw(1)}=\alpha\times[\tau/2\tau_{sw(1)}]\simeq0.42\pm0.07$. Additionally, the comparison of the curves' slopes in inset $(d)$ of Fig.~\ref{fig:fig2} indicates that $v_1(1)\sim v_2(k)$, where $v_n(k)$ is the wave velocity of pulse $k$ in chain $n$. Thus remembering that $\tau_{sw(n)}(k)=R_n\sqrt{10}/v_n(k)$, the ratio of characteristic durations $\tau_{sw(1)}/\tau_{sw(2)}\sim R_1/R_2\sim2$ leads to $\tau_*/2\tau_{sw(2)}\simeq0.84\pm0.14$. Consequently, the duration between two successive secondary collisions of the striker equals approximately the duration of one single transmitted SW, as stated above. However, the elapsed duration $\tau_*$ is slightly amplitude dependent, through $v_n(k)\propto F_m^{1/6}(k)$, and slowly increases as the force at interface decreases. Thus, the exponential approximation should be considered with care and may remain valid only if considered over a limited range of force amplitude at the interface, i.e. over a small number of pulses in the SWT.

\section{Dynamical simulations and quantitative analysis by means of quasi-particle description}\label{sec:robust}

\begin{figure}[top]
\centering\includegraphics[width=0.45\textwidth]{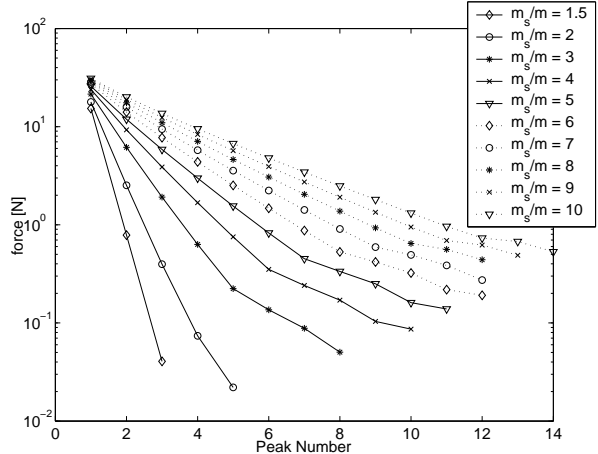}
\caption{Numerical simulations. The figure shows force peaks recorded in a bead in a granular chain where the bead is sufficiently far from the edge. The force peaks are recorded as a function of the peak number. The force decay is found to be approximately exponential in this calculation where the study has been carried out via direct solution of the Newtonian equations for the Hertzian grains.}\label{fig:fig4}
\end{figure}

Dynamical simulations have been performed, in which we have allowed strikers of different masses to hit the edge of a sufficiently long chain. The equations of motion have been solved via the velocity Verlet algorithm. Dissipative effects have been ignored in our numerical study although such effects can be readily incorporated.  One finds that sufficiently far from the chain edge the force felt by an individual grain can be plotted as a function of time for each value of the ratio $m_s/m$. From the conclusions drawn in last section, it turns out that it appears better to plot the force felt as a function of the force peak sequence instead of the time. This data is shown in Fig.~\ref{fig:fig4}. 

\subsection{Single solitary wave energy and momentum: an equivalent quasi-particle description}\label{sec:robust_sw}

We present here a complementary model to describe the interaction of a striker with an entire chain, based on energy and momentum conservation. Before the collision with the chain, the striker carries energy and momentum, $E_s=(1/2)m_sV_s^2$ and $Q_s=m_sV_s$, respectively. The impact generates a nonlinear wave in the chain, and one knows that the hertzian potential energy stored at contact $n$ is $U_n=(2\kappa/5)\delta_n^{5/2}$, the kinetic energy of a bead is $K_n=(1/2)mV_n^2$ and its momentum is $Q_n=mV_n$. Assuming that a single solitary wave propagates in the chain, one can estimate total energy $E_{sw}=\sum{(K_n+U_n)}$ and momentum $Q_{sw}=\sum{Q_n}$ by summing over all the contacts and all the beads. Denoting $Q_n/(2R)$, $K_n/(2R)$ and $U_n/(2R)$, as momentum, kinetic and potential energies per unit length respectively, the total energy and momentum in the chain can also be determined by integrating these quantities over the entire chain length. The kinetic energy $K_{sw}$, the potential energy $U_{sw}$, the total energy $E_{sw}$, and the momentum $Q_{sw}$ of a single SW sufficiently far from the edges are,
\begin{eqnarray}\label{eq:energy_momentum}
K_{sw} &=& \frac{mv^2}{4R}\int{\Psi^2dx} = \left[\frac{(2R)^3\Psi_m^{5/2}}{\theta\sqrt{10}}\right]W_8,\label{eq:sw_kin_nrj}\\
U_{sw} &=& \frac{2\kappa(2R)^{3/2}}{5}\int{\Psi^{5/2}dx} = \left[\frac{(2R)^3\Psi_m^{5/2}}{\theta\sqrt{10}}\right]W_{10},\label{eq:sw_pot_nrj}\\
E_{sw} &=& K_{sw}+U_{sw} = \left[\frac{(2R)^3\Psi_m^{5/2}}{\theta\sqrt{10}}\right](W_8+W_{10}),\label{eq:sw_tot_nrj}\\
Q_{sw} &=& \frac{mv}{2R}\int{\Psi dx} = \left[\frac{(2R)^3\Psi_m^{5/4}}{\sqrt{3\theta/\pi\rho}}\right]W_4\label{eq:sw_mom}
\end{eqnarray}
where $W_n=\int_{0}^{\pi/2}{\cos^n{(\xi)}d\xi}$ is the Wallis integral\footnote{http://fr.wikipedia.org/wiki/Int\'egrales\_de\_Wallis}, which is such that $W_{n\geq2}=[(n-1)/n]W_{n-2}$ and $W_0=\pi/2$.\\

In order to check the accuracy of previous estimations, one can compare for instance the kinetic to potential energy ratio, $K_{sw}/U_{sw}=W_{8}/W_{10}=10/9$, to the Virial theorem prediction, which states that $K_{sw}/U_{sw}=5/4$ if inter-grains forces derive from the Hertz potential. The difference remains weak (less than $10$\%) and is consistent with the long wavelength approximation to obtain continuous equation Eq.~\ref{eq:strain_equation} and solution Eq.~\ref{eq:strain_solution}. This difference could be lowered by calculating higher order correction terms of the potential energy from the Lagrangian density of Eq.~\ref{eq:strain_equation}~\cite{Nesterenko2001}.

Interestingly, the proportionality of kinetic and potential energies and the spatial compactness of a single SW give an opportunity to describe its dynamics as a quasi-particle~\cite{Nesterenko1994,Nesterenko1995,Nesterenko2001,Daraio2004}. Defining an equivalent particle of mass $m_{sw}$ moving at velocity $V_{sw}$, such that SW momentum and energy are $Q_{sw}=m_{sw}V_{sw}$ and $E_{sw}=(1/2)m_{sw}V_{sw}^2$, one finds,
\begin{eqnarray}
m_{sw} = \frac{Q_{sw}^2}{2E_{sw}} &=& \Omega\times m \mbox{ where } \Omega=\frac{W_4^2\sqrt{10}}{W_8+W_{10}},\label{eq:sw_mass}\\
V_{sw} &=& \Gamma\times V_m \mbox{ where } \Gamma=\frac{W_8+W_{10}}{W_4},\label{eq:sw_velocity}
\end{eqnarray}
where $V_m=v\times\Psi_m$ is the maximum velocity of beads in the chain, and where $\Omega\simeq1.345$ and $\Gamma\simeq1.385$. The effective mass agrees with Daraio's and Nesterenko's prediction~\cite{Daraio2004}, who reported a SW equivalent mass about $1.4$ times the mass of a single bead of the chain. The significance of the SW effective mass may be understood by considering that the strain wave spatial extent is a few beads wide, and $m_{sw}$ roughly corresponds to the total mass of these beads weighted by the wave envelope (e.g. by their own velocities). This effective model may constitute a simple and helpful tool to predict SW energy and momentum transmissions, by using hard sphere binary collisions formalism. As a first example, one can predict that a striker with mass $m_s=m$ and initial velocity $V_s$ impacting an uncompressed monodisperse chain made of bead with mass $m$ generates a SW with maximum velocity $V_m/V_s=2/[(1+\Omega)\Gamma]\simeq0.616$. When the SW further reaches the end of the chain, the last bead is ejected with velocity $V_l/V_s=4\Omega/(1+\Omega)^2\simeq0.978$. Previous results are both in satisfactory agreement with Chatterjee's numerical results~\cite{Chatterjee1999}, who reported that $V_m/V_s\simeq0.682$ and $V_l/V_s\simeq0.986$. As a second example, Nesterenko showed that a single SW propagating from a chain of light beads to a chain of heavy ones generates single reflected and transmitted SWs at the interface~\cite{Nesterenko1994}. Applying conservation laws on effective quasi-particles' momentum and energy gave reflected and transmitted amplitudes estimates in agreement with computer calculations~\cite{Nesterenko1994,Nesterenko1995}. To the best of our knowledge, quantitative description of single SW propagating from a chain of heavy beads to a chain of light ones has not been yet provided. This is presented in following subsection.

\subsection{Solitary wave train energy and momentum}\label{sec:robust_swt}

We estimate here energy and momentum of a SWT. We assume first that a SWT is composed of a sufficiently large number of well defined and separated single SWs, and second that their amplitudes, expressed in terms of the strain, decrease exponentially, $\Psi_m(k)\propto\exp{(-\beta k)}$, where $\Psi_m(k)$ denotes the amplitude of the $k^{th}$ solitary wave, and $\beta\in\Re^+$. Total energy and momentum of the chain are thus,
\begin{eqnarray}
E_{swt} &=& \sum_{k=1}^{\infty}{E_{sw}(k)}=E_{sw}(1)/[1-\exp{(-5\beta/2)}],\label{eq:swt_nrj}\\
Q_{swt} &=& \sum_{k=1}^{\infty}{Q_{sw}(k)}=Q_{sw}(1)/[1-\exp{(-5\beta/4)}].\label{eq:swt_mom}
\end{eqnarray}

Considering that energy and momentum are totally transmitted ($E_{swt}=E_s$ and $Q_{swt}=Q_s$) at long time if the mass ratio $(m_s/m)$ is large~\cite{Lazaridi1985,Nesterenko1994,Nesterenko1995}, one finds the following relation,
\begin{equation}\label{eq:swt_mass}
m_s = \frac{Q_s^2}{2E_s} = \frac{Q_{swt}^2}{2E_{swt}} = \frac{\Omega\times m}{\tanh{(5\beta/8)}}.
\end{equation}

\begin{figure}[top]
\centering\includegraphics{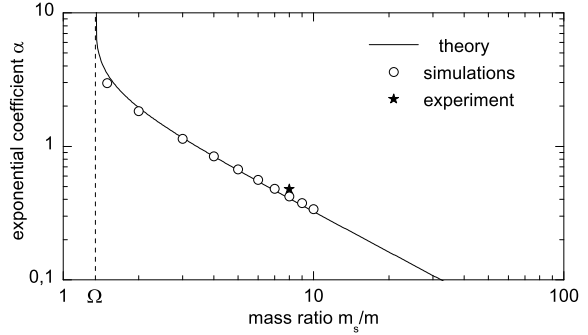}
\caption{Coefficient $\alpha$, defined such that $F_m(k)\propto\exp{(-\alpha k)}$ where $F_m(k)$ is the maximum force of the $k^{th}$ transmitted SW, when a striker of mass $m_s$ collides a monodisperse chain of beads with mass $m$. Theory is obtained from Eq.~\ref{eq:alpha_robust} and simulations data comes from Fig.~\ref{fig:fig4} and~\cite{Sokolow2007}. The experimental point corresponds to measurements done in the stepped chain described in Fig.~\ref{fig:fig2}, with $m_s/m=m_1/m_2=8$. Experimental coefficient is $\alpha_{exp}=0.48\pm0.03$ and theory predicts $\alpha_{th}=0.41$.}\label{fig:fig5}
\end{figure}

Using the relation between force and strain, $F_m(k)\propto\Psi_m^{3/2}(k)\propto\exp{(-\alpha k)}$, so that $\alpha=3\beta/2$, one finally finds,
\begin{equation}\label{eq:alpha_robust}
\alpha \equiv \left(\frac{6}{5}\right)\ln{\left[\frac{1+(m/m_s)\Omega}{1-(m/m_s)\Omega}\right]},
\end{equation}
where it can be seen that $\alpha\in\Re^+$ implies that previous developments and results are physically valid if $(m_s/m)>\Omega$. The comparison between Eq.~\ref{eq:alpha_robust}, simulations from Fig.~\ref{fig:fig4} and~\cite{Sokolow2007}, and the experiment from Fig.~\ref{fig:fig2}, is shown in Fig.~\ref{fig:fig5}. A satisfactory agreement is shown although numerical and experimental SWTs seem to contain a finite number of single SWs. A comparison between the expression given by Eq.~\ref{eq:alpha_robust} and the approximate result given by Eq.~\ref{eq:alpha_simple} confirms that the qualitative description given in section~\ref{sec:qualitative} is meaningful, provided the interaction between the striker and the chain is considered as an interaction between the striker with a SW equivalent quasi-particle of mass $m_{sw}=\Omega\times m$, rather than with the first bead of mass $m$. In addition, the result presented in Eq.~\ref{eq:alpha_robust} depicts experimental and numerical observations in the sense that the coefficient $\alpha$ does not depend on incident SW amplitude~\cite{Sokolow2007}.

The analysis above, in addition, sheds some light on the shock absorption capabilities of linear chains. Indeed, as shown experimentally in Fig.~\ref{fig:fig3}, the normalized force at the interface of the stepped chain also appeared independent of incident SW amplitude. According to Eqs.~\ref{eq:swt_nrj}-\ref{eq:alpha_robust} (when $m_s\geq\Omega\times m$), the leading transmitted SW ($k=1$) energy and momentum are indeed found proportional to incident energy and momentum, and respective coefficients of proportionality only depend on mass ratio,
\begin{eqnarray}
E_{sw}(1)&=&\frac{4\Omega(m/m_s)}{[1+\Omega(m/m_s)]^2}\times E_s,\label{eq:blast_protection_e}\\
Q_{sw}(1)&=&\frac{2\Omega(m/m_s)}{1+\Omega(m/m_s)}\times Q_s.\label{eq:blast_protection_q}
\end{eqnarray}

These last results underline the specific and highly nonlinear behavior of non-loaded granular chains, since the larger the striker is, the longer the duration of the SWT (i.e. the smaller $\alpha$ is) and the smaller the leading transmitted SW's energy and momentum are. This, in particular, might be of interest to design efficient blast protection systems to prevent from the impact of large objects.

\section{Conclusion}\label{sec:conclusion}

In summary, we have presented a detailed study of the formation of solitary wave trains in a non-loaded chain of beads. In accordance with existing results, the qualitative description demonstrated that the impact a light striker on an alignment of spheres generates a single solitary wave, while the impact of a large striker induces a solitary wave train. The experimental study, the detailed theoretical analysis, and the dynamical simulations using Newtonian equations revealed that the force peaks felt by any grain in the chain decays approximately exponentially with the peak sequence, as a solitary wave train passed through it. The use of an equivalent quasi-particle description, in addition to energy and momentum conservation, allowed finding a simple relationship between the shape of the solitary wave train and the striker/bead mass ratio. The prediction is in satisfactory agreement with simulations and experiment. Provided the striker is slightly heavier than a single bead, the overall strength of the transmitted wave lowers and its total duration rises as the striker mass is increased. This nontrivial and unique feature should prove valuable when optimizing blast protection capabilities of such a system. For instance, given the time and amplitude scale of the impulse force, our results help to determine the physical dimensions of a bead protection system.

\begin{acknowledgements}
FM and SJ acknowledge support of Conicyt under Fondap Program $N^{o}$ 11980002. SS and AS acknowledge support of US Army Research Office.
\end{acknowledgements}

\bibliographystyle{unsrt}
\bibliography{SWT_Job}

\end{document}